\documentclass[doc]{apa6}
\usepackage[T1]{fontenc}
\usepackage[latin9]{inputenc}
\usepackage{array}
\usepackage{rotating}
\usepackage{multirow}
\usepackage{adjustbox}
\usepackage{siunitx}
\usepackage{graphicx}
\usepackage{subcaption}
\usepackage{amsmath,amssymb}
\usepackage{mathrsfs}
\usepackage{textgreek}
\usepackage{fixltx2e}
\usepackage{bm}
\usepackage{url}
\usepackage[english]{babel}

\usepackage[
backend=biber,
style=authoryear,
citestyle=apa
]{biblatex}

\makeatletter
\title{Oscillatory dynamics in complex recurrent neural networks}
\shorttitle{Oscillatory dynamics in RNN}

\twoauthors{Rakesh Sengupta}{P. V. Raja Shekar}
\twoaffiliations{Center for Creative Cognition, SR University, Warangal, India}{Center for Creative Cognition, SR University, Warangal, India}

\abstract{Spontaneous oscillations measured by Local field potentials (LFPs), electroencephalograms and magnetoencephalograms 
	exhibits variety of oscillations spanning frequency band ($1-100$ Hz) in animals and humans. 
	Both instantaneous power and phase of these ongoing oscillations have commonly been observed to correlate with pre-stimulus processing 
	in animals and humans. However, despite of numerous attempts it is not fully clear whether the same mechanisms can give 
	rise to a range of oscillations as observed in vivo during resting state spontaneous oscillatory activity of the brain. In the current paper we show how oscillatory activity can arise out of 
	general recurrent on-center off-surround neural network. The current work shows (a) a complex valued
	input to a class of biologically inspired recurrent neural networks can be shown to be mathematically
	equivalent to a combination of real-valued recurrent network with real-valued feed forward network, (b) such a 
	network can give rise to oscillatory signatures. We also validate the conjecture with results of simulation 
	of complex valued additive recurrent neural network. }

\begin{document}
\maketitle
\section{Introduction}
It has always been a standard practice to break down the brain signals
obtained in electroencephalography (EEG) or magnetoencephalography
(MEG) in terms of their frequency components obeying the laws of Fourier
decomposition. It is even commonplace to compartmentalize the frequencies
into neat bunches like $\theta$ (4-8 Hz), $\alpha$ (8-12 Hz), etc.
However, the important question is whether these bands are just a
matter of convenience or is there an underlying reality to the oscillatory
model for the brain. Some recent work has tried to tie the oscillatory
framework to underlying spiking model for neurons \parencite{Deco2009,Nakagawa2014}.
This question can be solved only by looking at mathematical properties
of neural codes. \\

In recent years the understanding of neural codes has provided us
with insights that go beyond the concepts of rate coding and it is
increasingly more commonplace to speak of temporal codes that use
spike timing and phase information in order to transmit and process
information reliably \parencite{Stanley2013,VanRullen2001,Masquelier2012,Gautrais1998}.
However, most of these attempts are to look at neural codes after
the presentation of stimulus. Recently, some studies have looked at
pre-stimulius brain states in MEG and EEG based studies and have found
that it is possible to predict conscious detection of stimuli based
on pre-stimulus oscillatory brain activity \parencite{Mathewson2009,Weisz2014,May2012,Keil2012}.
For instance in case of near threshold stimuli some researchers have
found the pre-stimulus $\alpha$ frequency band modulation to be important
\parencite{Weisz2014}. These attempts have drawn a large amount of interest,
but have revealed little towards a theoretical or physiological understanding
of such phenomena. In the current work we have tried to start from
a minimal number of assumptions regarding a neuronal assembly and
tried to show how it is possible to theoretically derive such an oscillatory
dynamics. \\

\section{Methods}

% For figure citations, please use "Fig." instead of "Figure".
%\begin{description}
\paragraph*{Conjecture 1} If a neural assembly $\mathbb{S}$ of $N$ neurons
consists of both feed-forward and recurrent connections with a finite
variable bound refractory period $\tau$ between the feed-forward
and recurrent connections, the equilibrium solution for the assembly
can be characterized by a class of oscillatory functions.
\paragraph*{Justification} If a neural assembly comprises of $n_{1}$feed-forward
neurons and $n_{2}$ recurrent neurons ($n_{1}+n_{2}\leq N)$, then
the general activations of the pools of neurons will be given by 
\begin{equation}
	\mathbf{\dot{x}_{1}}=\frac{d\mathbf{x_{1}}}{dt}\}_{n_{1}}=\textrm{sgn}(\xi^{T}\xi\mathbf{x_{1}}-\theta)
	\label{eq:n1FirstOrder}
\end{equation}
\begin{equation}
	\mathbf{\dot{x}_{2}}=\frac{d\mathbf{x_{2}}}{dt}\}_{n_{2}}=-\mathbf{x_{2}}+\mathbf{F}.\mathbf{x_{2}}+
	\eta\mathbf{I}_{\mathbf{x_{1}}}\label{eq:n2FirstOrder}
\end{equation}
where $\xi$ is the weight matrix for the feed-forward connections,
$\theta$ is the threshold, $\eta$ is a scaling multiplier, $\mathbf{F}$
is the non-linear transfer function and $\mathbf{I}_{\mathbf{x_{1}}}$
is the input coming from the feed-forward networks\footnote{If we consider a general recurrent shunting
	networks with dynamics given by 
	\begin{equation}
		\dot{x_i}=-A_i x_{i}+(B_i -x_{i})(I_i + \mathcal{S}(x_i ))-(x_i + C_i) \left(J_i + \sum_{j\neq i} w_{ji} \mathcal{S}(x_j)
		\right)
		\label{eq:shuntingFirstOrder}
	\end{equation}
	where $I$ and $J$ are excitatory and inhibitory inputs and $\mathcal{S}$ is a sigmoid function, we can transform it 
	with simple variable change ($y_i = x_i + C_i$) to the form
	\begin{equation}
		\dot{y_i}=y_i \left( b_i (y_i ) - \sum_{j=1}^{n} w_{ji} \mathcal{S} (y_j - C_j ) \right)
	\end{equation}
	where,
	\begin{equation}
		b_i (y_i) = \frac{1}{y_i} \left\{A_i C_i - (A_i + J_i)y_i + (B_i +C_i -y_i)\left[I_i + \mathcal{S}(y_i -C_i )
		\right]\right\}
	\end{equation}
	It can be shown that in absence of input and near equilibrium, second order dynamics
	of this network will be very similar to Eq. \ref{eq:x2dotdot2} with an extra term of the order of $y_i^2$,
	which does not change the general conclusions of the paper.
	
}. Now, for the network
to be useful there needs to be sustained activity (provided by the
recurrent network, see \parencite{sengupta1}) as well as the ability
to restart the network dynamics. The latter is provided by assuming
that a smoothing regularizer is provided which varies inversely with
the network output \parencite{Wu1996} (to smooth the network against
noisy perturbations), as well as a global decay parameter. In such
a network, the overall output function $\overline{<\mathbf{x_{1}}+\mathbf{x_{2}}>}$
(average expectation value of the network output constructed in the
line of mean activation in \parencite{sengupta1}), varies between 0
and $\overline{<\mathbf{x_{2}}>}+I$, where $I$ is the average normalized
input to the network. This alone is sufficient to show the possibility
of oscillatory solutions. However, to be more rigorous, let us look
at the second order time evolution for the feed-forward and recurrent
populations. From Eq. \ref{eq:n1FirstOrder} we get.
\begin{equation}
	\mathbf{\ddot{x}_{1}}=\frac{d^{2}\mathbf{x_{1}}}{dt^{2}}\}_{n_{1}}=\pm2\delta(\xi^{T}\xi\mathbf{x_{1}}-\theta)\label{eq:n1SecondOrder}
\end{equation}
Here $\delta(x)$ is the Dirac delta function. From Eq. \ref{eq:n2FirstOrder}
we have, 
\begin{eqnarray}
	d\mathbf{\dot{x}_{2}} & = & -d\mathbf{x_{2}}+d(\mathbf{F}.\mathbf{x_{2}})\label{eq:dx2dot}\\
	&  & =-d\mathbf{x_{2}}+(d\mathbf{F}).\mathbf{x_{2}}+\mathbf{F}.d\mathbf{x_{2}}\label{eq:dx2dot2}
\end{eqnarray}
If $\mathbf{F}$ is a smooth transfer function then the third term
in the right hand side of Eq. \ref{eq:dx2dot2} becomes negligible.
Thus we have 
\begin{equation}
	\mathbf{\ddot{x}_{2}}=-\mathbf{\dot{x}_{2}}+\mathbf{\dot{F}}.\mathbf{x_{2}}\label{eq:x2dotdot}
\end{equation}
\begin{equation}
	\mathbf{\ddot{x}_{2}}=(1-\mathbf{F})\mathbf{x_{2}}+\mathbf{\dot{F}}\mathbf{x_{2}}+\mathcal{\mathcal{O}}(\mathbf{I}_{\mathbf{x_{1}}})\label{eq:x2dotdot2}
\end{equation}
For the recurrent network, near equilibrium (here we assume absence
of input because of the finite refractory period of $\tau$), the
nonlinear operators $\mathbf{F}$ and $\mathbf{\dot{F}}$ become quasi-linear
operators, and thus we have 
\begin{equation}
	\mathbf{\ddot{x}_{2}}=\frac{d^{2}\mathbf{x_{2}}}{dt^{2}}\}_{n_{2}}=-\parallel\mathbf{F}-(1+\mathbf{\dot{F}})\parallel\mathbf{x_{2}}\label{eq:n2SecondOrder}
\end{equation}
This has a form of simple eigenvalue problem. The network will have
oscillatory solution if the operator $\parallel\mathbf{F}-(1+\mathbf{\dot{F}})\parallel$
is positive Hermitian. A suitable example is the leaky accumulator
network described in \parencite{sengupta1} where the matrix representing
the operator becomes real symmetric. Thus near equilibrium, considering
$\mathbf{F}$ is a quasi-linear approximation $<\mathbf{x_{2}}>$,
the mean state of the neural assembly near equilibrium / steady state
will have a periodic solution $\xi_{n}=Ae^{i(\omega_{n}t-\theta)}$
with $\mathbf{x_{1}}$ contributing to the phase $\theta$. The full
neural field dynamics $\psi(x,t)$ can be constructed along the lines
of \cite{Jirsa1996} with both spatial and temporal components taken
into account as the following 
\begin{equation}
	\psi(x,t)=\sum\xi_{n}(t)\exp(inkx)\label{eq:spatio_temporal_wave}
\end{equation}
Thus the predictions from this analytical framework can be easily ported
to neural mass models as well.

\paragraph*{Corollary} An assembly of recurrent neurons with complex-valued
inputs and outputs, is formally equivalent to a neural assembly of
independent feed-forward and recurrent neurons.
\paragraph*{Justification} Let us start with Cohen-Grossberg generalized networks
of additive variant with a nonlinear activation function, like the
network described by \cite{sengupta1,usher,bogacz}. If inputs and
outputs to the network are given by a vector of complex numbers $\bar{\mathbf{z}}_{\{1\times m\}}=\bar{\mathbf{x}}_{\{1\times m\}}+i\bar{\mathbf{y}}_{\{1\times m\}}$
for a network of $m$ nodes, the network dynamics is governed by
\begin{equation}
	\frac{dz_{j}}{dt}=-z_{j}+c_{1}F(z_{j})-c_{2}\sum_{k\neq j}F(z_{k})+I_{j}+\mathrm{noise}\label{eq:complexadditive}
\end{equation}
where $F(z)=z/(1+z)$. Ignoring noise for the time being, if we decompose
the real and imaginary parts of \ref{eq:complexadditive}, we have
\begin{equation}
	\begin{split}
		\frac{dz_{j}}{dt}
		&=-z_{j}+c_{1}\frac{z_{j}+\left|z_{j}\right|^{2}}{1+\left|z_{j}\right|^{2}+2\mathrm{Re}(z_{j})}\\
		&-c_{2}\sum_{k\neq j}\frac{z_{k}+\left|z_{k}\right|^{2}}{1+\left|z_{k}\right|^{2}+2\mathrm{Re}(z_{k})}+I_{j}\label{eq:complexAlgebra}
	\end{split}
\end{equation}
And thus the separated real and imaginary parts yield two equations
given by, 
\begin{equation}
	\begin{split}
		\frac{d\mathrm{Re}(z_{j})}{dt}
		&=-\mathrm{Re}(z_{j})+c_{1}\frac{\mathrm{Re}(z_{j})}{1+\left|z_{j}\right|^{2}+2\mathrm{Re}(z_{j})}\\
		&-c_{1}\sum_{k\neq j}\frac{\mathrm{Re}(z_{k})}{1+\left|z_{k}\right|^{2}+2\mathrm{Re}(z_{k})}+\mathrm{Re}(I_{j})+\mathcal{O}(\left|z\right|^{2})\label{eq:realX}
	\end{split}
\end{equation}
\begin{equation}
	\frac{d\mathrm{Im}(z_{j})}{dt}=\sum c'_{jk}\mathrm{Im}(z_{k})+\mathrm{Im}(I_{j})\label{eq:imagY}
\end{equation}
where $c'_{jk}$ are constants. Making suitable substitutions ($x_{j}\leftarrow2(\mathrm{Re}(z_{j})+\left|z_{j}\right|^{2}/2)$,
and $y_{j}\leftarrow\mathrm{Im}(z_{j})$ ), we have
\begin{equation}
	\frac{dx_{j}}{dt}=-x_{j}+c"_{1}F(x_{j})-c"_{2}\sum_{k\neq j}F(x_{k})+I_{j}+\textrm{constant terms}\label{eq:recurrentReal}
\end{equation}
\begin{equation}
	\frac{dy_{j}}{dt}=\sum c'_{jk}y_{k}+I'_{j}\label{eq:feedforwardIm}
\end{equation}
From Eq. \ref{eq:recurrentReal} and \ref{eq:feedforwardIm} it is
clear that a combination of recurrent and feedforward networks can
function in a way to handle complex inputs and outputs and thus at
least the reverse of Conjecture 1 is true in certain cases\footnote{
	The above analysis also holds for a general sigmoid activation function. If $\sigma (x) = \frac{1}{1+e^{-x}}$,
	then from simple function approximation of a general sigmoidal function $f(x) = \sum_{i=1}^n c_i \sigma (x-a_i )$
	can be simply transformed in the complex domain as $f(x) \rightarrow \sum_{i=1}^n \frac{zc_i}{z+\alpha_i}$, where
	$\alpha_i = e^{a_i}$ (see Appendix C of \cite{mandic2001recurrent}). Thus the results of conjecture 2 are quite general.
}.\\

In the following section we show simulations involving additive neural networks giving rise to oscillatory
dynamics under complex inputs.\\

\section{Results}

In the above, we have shown the possibility of oscillatory brain states under certain equilibrium conditions
for a neural assemble consisting of both feed-forward and
recurrent connections. We simulated a recurrent on-center off-surround neural network governed by the dynamical equation
(complex version of the neural network used in \cite{sengupta1})

\begin{equation}
	\frac{dz_{i}}{dt}=-x_{i}+\alpha F(z_{i})-\beta\sum_{j=1,j\neq i}^{N}F(z_{j})+I_{i}+noise\label{eq:model_add}
\end{equation}

where $z_i$ are complex valued activations of the nodes indexed 1 to $N$ (we used 10 nodes for the simulation). All the nodes are fully connected with self-excitation $\alpha$ (we used values 0.05, 0.55, 1.05, 1.55, 2.05, 2.55, 3.05, 3.55,
4.05, 4.55, 5.05) and lateral inhibition $\beta$ (we used 0.03, 0.08, 0.13, 0.18, 0.23, 0.28). Transfer function $F$ is defined as 

\begin{equation}
	F(z)=\frac{z}{1+z} \label{eq:activation_function}
\end{equation}

$I_i$ is a transient complex input for 200 ms steps with values clamped at ($0.7+0.1i$). \\

\begin{figure}
	\includegraphics[width=\columnwidth]{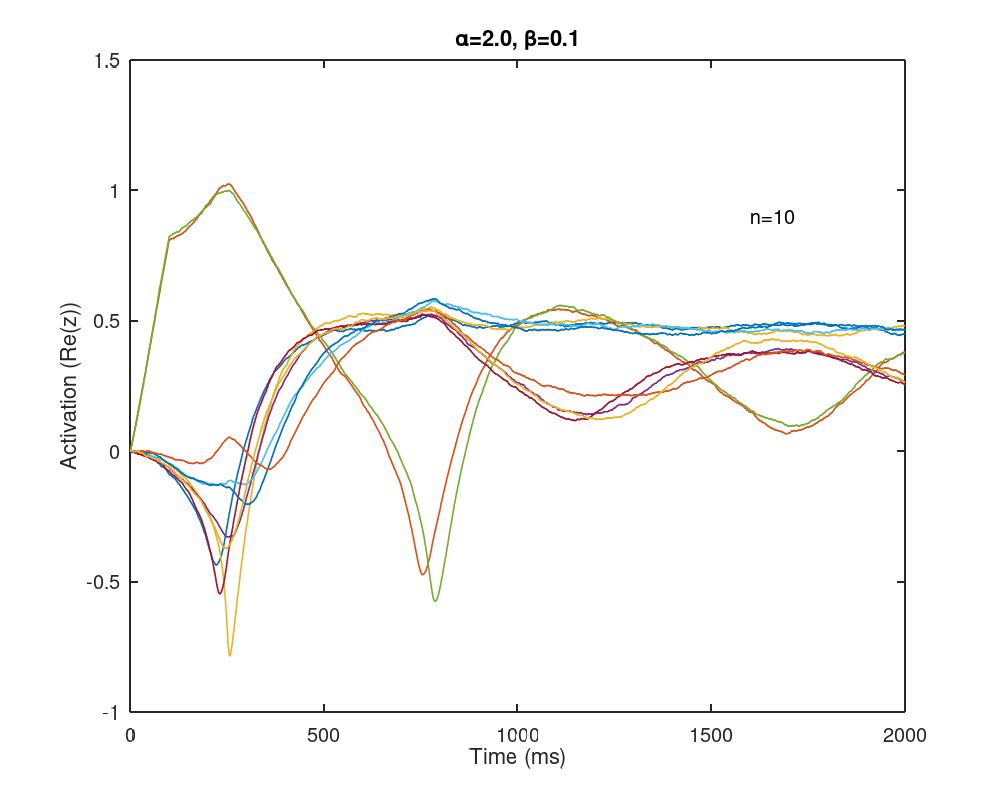}
	
	\caption{Real values of the activation for the nodes of the network where $\alpha = 2.0$ and $\beta = 0.1$.}

	\label{fig:osc1}
\end{figure}

We ran the numerical simulation for total 2000 ms steps using time steps of 0.01 using Euler method. Noise was sampled from a Gaussian distribution with mean 0 and standard deviation of 0.05. Fig. \ref{fig:osc1} depicts $\mathrm{Re}(z)$ for the 
network for network parameters  $\alpha = 2.0$ and $\beta = 0.1$. \\

\begin{figure}
	\includegraphics[width=\columnwidth]{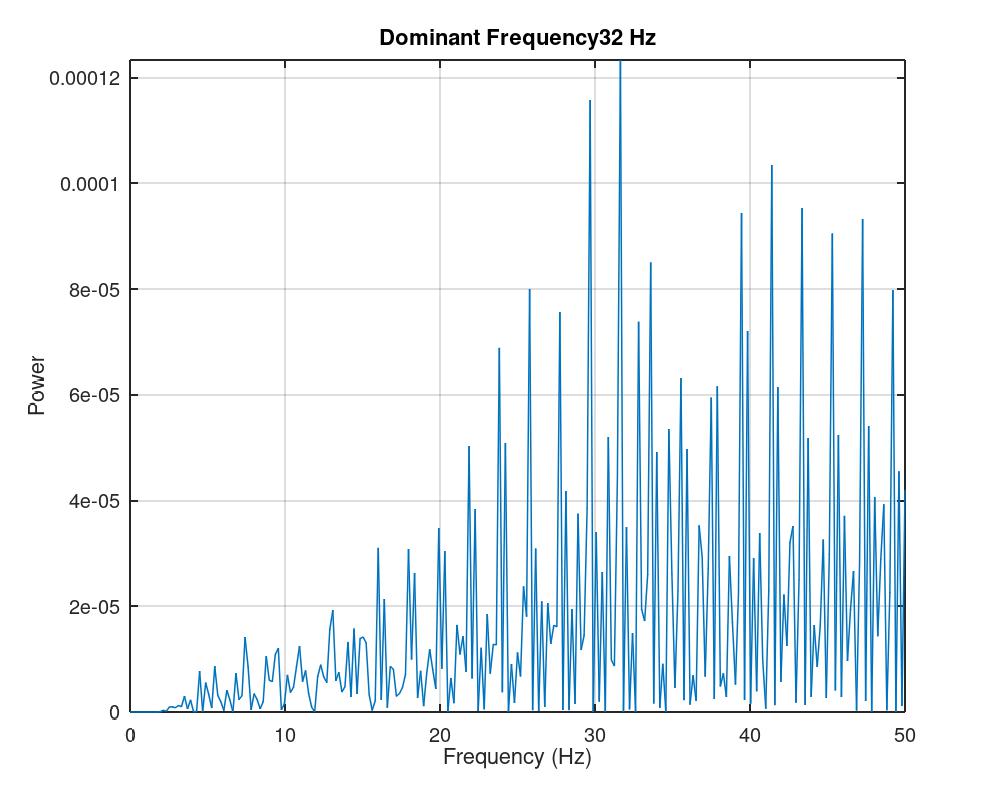}
	
	\caption{Frequency power analysis for $\mathrm{Re}(z)$ for the nodes of the network where $\alpha = 2.0$ and $\beta = 0.1$.}

	\label{fig:osc2}
\end{figure}

\begin{figure}
	\includegraphics[width=\columnwidth]{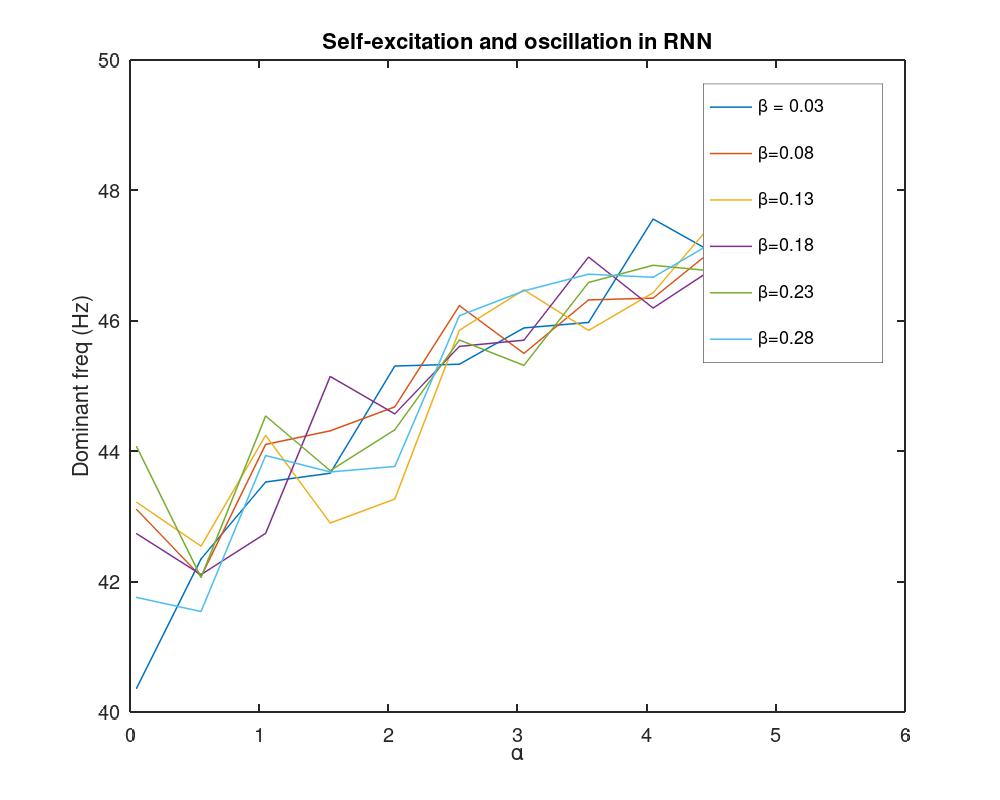}
	
	\caption{Average dominant frequency values for all the $\alpha$ and $\beta$ values in simulation. }

	\label{fig:osc3}
\end{figure}

We ran time-frequency analysis on $\mathrm{Re}(z)$ and the results are shown in Fig. \ref{fig:osc2}. We ran
100 simulations for each values of $\alpha$ and $\beta$ and Fig. \ref{fig:osc3} shows the average dominant frequency values
for all $\alpha$ and $\beta$. \\

\section{Discussion}

In the above, we have shown the possibility of oscillatory brain states under certain equilibrium conditions
for a neural assemble consisting of both feed-forward and
recurrent connections.
We also showed with help of simulations (Fig. \ref{fig:osc1}, \ref{fig:osc2} \& \ref{fig:osc3}) that simple
fully connected additive recurrent neural network with complex inputs can lead to oscillatory dynamics. We saw in Fig. \ref{fig:osc3}
that dominant frequency of the steady-state dynamics varies linearly with the self-excitation parameter ($\alpha$) of
the network, but not with the lateral inhibition ($\beta$). In the future we intend to investigate similar dynamics for other recurrent network variants like multiplicative and self-organized maps, etc.

\begin{appendix}
\section{Cognitive dynamics and pre-stimulus oscillations}

\paragraph*{Conjecture} If a neural assembly $\mathbb{S}$ of $N$ neurons
consists of both feed-forward and recurrent connections with a finite
variable bound refractory period $\tau$ between the feed-forward
and recurrent connections, the pre-stimulus brain states are determined
by the delay between the regions represented by the neural assemblies.

\paragraph*{Justification} The conjecture 1
has some interesting consequences. Firstly, it allows for brain states%
\footnote{Here we are using the term brain state to mean the state of the neural
	assembly involved in a particular task or function, not the entire
	brain. %
} to be defined in terms of a frequency or a distribution of frequencies.
In a case of a state with a distribution of frequencies we can think
of a characteristic frequency range representing the state, the characteristic
frequency being the one with maximal power. The conjecture also allows
us to think of the brain states as phenomenal superposition of oscillatory
dynamics, allowing us to deal with problems such as stimulus related
perturbations to brain states more efficiently.

The prevailing additive idea of brain states in the neuroimaging literature
needs no introduction. However, we will formally spell out the bare
essentials. Before a test condition appears to a subject the family
of brain states or the neural assembly $\mathbb{S}$ in question ($S$)
can be thought of as its resting state.
In the test condition, a new perturbation comes from our experimental
control ($S_{T}$). Or we can write, $S\leftarrow\bar{S}+S_{T}$. 

Now if the resting state is imagined as a standing wave, then we have from Eq. \ref{eq:spatio_temporal_wave}
$\bar{S} = \sum A\mathrm{cos}(kx)\mathrm{exp}(i\omega_0 t)$.
Since resting state can be taken to be not very location specific we can write
the rate of change of the state of neural assembly
given by $S$, when a perturbative brain state $S_{T}$ (generally
due to oncoming stimulus) interacts with the current state to be given
by 
\begin{equation}
	\frac{\partial S}{\partial t}=i\alpha \mathrm{exp}(\omega_0 t)+\frac{\partial S_{T}}{\partial t}\label{eq:brainEvolve}
\end{equation}
From the results of the previous section, if assume the brain states
and their perturbations to have solutions of the form $Ae^{i(\omega t-\theta)}$
we can write,
\begin{equation}
	a_{1}i\omega e^{i(\omega t-\theta)}=i\alpha e^{\omega_0 t}+a_{2}i\omega'e^{i(\omega't-\theta_{1})}\label{eq:condition}
\end{equation}
Here $\omega$ is the pre-stimulus brain state frequency that tries
to interact with the incoming perturbation characterized by $\omega'$.
Separating the real and the imaginary parts and applying the constraint
that the imaginary part must go to zero on the left and right hand
side of Eq. \ref{eq:condition} 
\begin{equation}
	a_{1}\omega\cos(\omega t-\theta)=\alpha \cos(\omega_0 t) + a_{2}\omega'\cos(\omega't-\theta_{1})\label{eq:condition2}
\end{equation}
Now considering the case $\omega\sim\omega'$, we have from Eq. \ref{eq:condition2}
(if $S$ and $S_{T}$ are in similar phase) we have for $t\rightarrow 0$, 
\begin{equation}
	(a_{1}-a_{2})\omega \cos(\omega t-\theta)= \alpha \label{eq:delay_condition}
\end{equation}

Thus we have our constraint,
\begin{equation}
	-1\leq \frac{\alpha}{(a_{1}-a_{2})\omega} \leq 1\label{eq:delay_condition2}
\end{equation}

Thus Eq. \ref{eq:delay_condition2} shows that the resonant frequency of
the pre-stimulus brain states in a region varies with $\alpha$\footnote{
	Now if we take the imaginary part of the Eq. \ref{eq:condition},
	then we have (for the case $\omega\sim\omega'$ if $S$ and $S_{T}$ are in similar phase)
	
	\begin{equation}
		(a_{1}-a_{2})\sin(\omega t-\theta)= 0\label{eq:condition3}
	\end{equation}
	
	\begin{equation}
		\omega t-\theta\sim n\pi
	\end{equation}
	\begin{equation}
		\mathcal{\mathcal{O}}(\omega t)\sim n\pi\label{eq:orderW}
	\end{equation}
	If $S$ and $S_{T}$ are anti-phase then also we have $\tan(\omega t-\theta)=\frac{a_{1}}{a_{2}}<\propto$
	and thus $\mathcal{\mathcal{O}}(\omega t)\sim\pi$.On the other hand
	for the condition $\omega\gg\omega'$, we have $\sin(\omega t-\theta)=0$,
	and thus Eq. \ref{eq:orderW} follows. The condition $\omega'\gg\omega$
	is not very informative as it assumed very high frequency stimulus
	for low frequency brain states. Thus overall we have that a pre-stimulus
	oscillatory brain state characterized by the frequency $\omega$,
	will be informative about the stimulus if the Eq. \ref{eq:orderW}
	holds. Interestingly, most brain dynamics is a time limited phenomena,
	and the frequency $\omega$ thus depends upon the delay between the connections
	as shown in \ref{eq:delay_condition2}. 
	However these delays are related to the biophysical process rates. Thus this formulation
	gives a very natural way of looking at which frequencies might appear
	in the dominant. For instance, if the delay $\tau=100$ ms, then $\nu=\frac{\omega}{2\pi}=\frac{1}{\tau}$
	would be of the order of 10 Hz. This idea is derived here in a non-trivial
	manner, i.e., just having a refractory period of 100 ms does not guarantee
	oscillatory solutions. Here we have shown how it is possible to have
	oscillations in the $\alpha$ band can be thought to arise from internal
	brain dynamics.}. Interestingly,
$\alpha$ is a spatial term dependent on the spatial connectivity as seen above.
Thus the frequencies generated in the oscillatory pre-stimulus brain
will depend upon the delay between the regions.

%\end{description}

Thus there are two consequences of the analytical framework. It shows how a general oscillatory activity can 
be generated in the particular brain region having both feedforward and recurrent connections.
Secondly it connects the general oscillatory signals in the brain that arises from connected
regions to be dependent upon the delay in the network connectivity arising from spatial factors.
\end{appendix}
%\bibliographystyle{apalike}
%\bibliography{references2}

\printbibliography

\end{document}